\documentclass[aps,prb,preprint,superscriptaddress]{revtex4-1}

\usepackage{graphicx}
\usepackage{bm}

\begin{document}

\title{Multiferroic properties of an \aa kermanite Sr$_2$CoSi$_2$O$_7$ single crystal in high magnetic fields}

\author{M. Akaki}
\email[]{m-akaki@sophia.ac.jp}
\affiliation{Department of Physics, Sophia University, Tokyo 102-8554, Japan}

\author{H. Iwamoto}
\affiliation{Department of Physics, Sophia University, Tokyo 102-8554, Japan}

\author{T. Kihara}
\affiliation{The Institute for Solid State Physics (ISSP), The University of Tokyo, Chiba 277-8581, Japan}

\author{M. Tokunaga}
\affiliation{The Institute for Solid State Physics (ISSP), The University of Tokyo, Chiba 277-8581, Japan}

\author{H. Kuwahara}
\affiliation{Department of Physics, Sophia University, Tokyo 102-8554, Japan}

\date{\today}

\begin{abstract}
The magnetic and dielectric properties of \aa kermanite Sr$_2$CoSi$_2$O$_7$ single crystals in high magnetic fields were investigated.
We have observed finite induced electric polarization along the $c$ axis in high fields, wherein all Co spins were forcibly aligned to the magnetic field direction. 
Existence of the induced polarization in the spin-polarized state accompanied with the finite slope in the magnetization curve suggests the possible role of the orbital angular momenta in the excited states as its microscopic origin.
The emergence of the field-induced polarization without particular magnetic order can be regarded as the magnetoelectric effects of the second order from the symmetry point of view.
A low magnetic field-driven electric polarization flip induced by a rotating field, even at room temperature, has been successfully demonstrated. 
\end{abstract}

\pacs{75.85.+t, 75.50.Ee, 77.80.-e}

\maketitle

Magnetoelectric multiferroic materials have attracted both experimental and theoretical interest from the point of view of fundamental science as well as for the possible applications of their giant magnetoelectric effects \cite{Fiebig,Tokura,revArima}.
In these materials, strong coupling or cross-correlation between magnetism and dielectricity is often realized through a magnetically-induced electric polarization, which is well explained by several microscopic mechanisms \cite{revArima}. 
The generally accepted model is the spin current mechanism \cite{Katsura}.
In materials with a cycloidal spin structure (e.g., TbMnO$_3$ \cite{Kimura}), electric polarization is induced by this spin current mechanism, which predicts the electric polarization ${\bm P}\propto \sum_{i,j} {\bm e}_{ij}\times ({\bm S}_i \times {\bm S}_j)$, with ${\bm S}_i$ and ${\bm S}_j$ being neighboring spins connected by a unit vector ${\bm e}_{ij}$.
Another origin of magnetically-induced ferroelectricity is the exchange striction mechanism.
In materials with multiple inequivalent magnetic sites, even a collinear spin structure may induce ferroelectricity via magnetostriction caused by the symmetric exchange interaction $({\bm S}_i \cdot {\bm S}_j)$ \cite{exArima}.
This type of ferroelectricity is realized in highly distorted orthorhombic $R$MnO$_3$ compounds ($R$=Y, Ho, ..., Lu) with $E$-type antiferromagnetism \cite{Ishiwata}, DyFeO$_3$ \cite{DyFeO3}, and Ca$_3$(Co,Mn)$_2$O$_6$ \cite{CoMn}.
Recently, a third mechanism of magnetically-induced electric polarization has been proposed, namely transition metal-ligand ({\it p-d}) hybridization, dependent upon the spin direction \cite{pdArima}.
The magnetoelectric properties of delafossite CuFeO$_2$ \cite{CuFeO2} and \aa kermanite Ba$_2$CoGe$_2$O$_7$ \cite{BaCoGe,Murakawa} are thought to be caused by this {\it p-d} hybridization mechanism.

Ever since an intriguing magnetoelectric effect was discovered in \aa kermanite Ca$_2$CoSi$_2$O$_7$ \cite{Akaki} and Ba$_2$CoGe$_2$O$_7$ \cite{BaCoGe,Murakawa}, the multiferroicity of these materials has been attracting much attention.
The present compound Sr$_2$CoSi$_2$O$_7$ has the same crystal structure as Ba$_2$CoGe$_2$O$_7$ with space group $P\overline{4}2_1 m$ at room temperature, and does not show a structural phase transition below 300~K\@.
As seen from the schematic crystal structure illustrated in Fig.~1(a), CoO$_4$ and SiO$_4$ tetrahedra are connected at their corners to form two-dimensional layers, with the layers stacking along the $c$ axis with intervening Sr layers.
The Sr$_2$CoSi$_2$O$_7$ compound shows a similar behavior to that of Ba$_2$CoGe$_2$O$_7$ based on magnetoelectric measurements in the absence of a magnetic field \cite{lt25,icm2009}. 
The Co magnetic moments have an easy plane anisotropy in the (001) plane with the $c$ axis being a hard axis, and show a staggered antiferromagnetic structure in the plane below $T_{\rm N}$=7~K\@.
In the antiferromagnetic phase, electric polarization without a poling electric field was induced by a magnetic field, with the value and direction of the induced electric polarization dependent upon the direction of the applied magnetic field.
Murakawa {\it et al.} reported the magnetoelectric effects of Ba$_2$CoGe$_2$O$_7$ in the canted antiferromagnetic state and interpreted the results in the framework of the {\it p-d} hybridization mechanism.
Contrary to the other microscopic scenarios, the electric polarization caused by this mechanism is irrelevant to the relative angle between the adjacent spins, and hence, emerges even in the fully spin-polarized state.
It is crucially important to investigate the magnetoelectric effect up to the field of spin saturation.
Therefore, we studied magnetoelectric effects in Sr$_2$CoSi$_2$O$_7$ in a wide range of magnetic fields and temperatures.

A single crystalline sample of Sr$_2$CoSi$_2$O$_7$ was grown using the floating zone method.
Through the x-ray diffraction measurements at room temperature, we confirmed that the sample had a tetragonal $P\overline{4}2_1 m$ structure without any impurity phase.
All specimens used were cut along the crystallographic principal axes into plate-like shapes by means of the x-ray back-reflection Laue technique.
The magnetization was measured using a commercial apparatus (Quantum Design, PPMS).
The magnetic field-induced electric polarizations were obtained by integration of the pyroelectric currents in the temperature scans at constant fields or the displacement currents in the magnetic field scans at constant temperatures.
Pulsed magnetic fields up to 55~T were generated with a duration time of 36~ms, using a nondestructive magnet in the International MegaGauss Science Laboratory of the ISSP at the University of Tokyo. The magnetization in pulsed magnetic fields was measured via the induction method, using coaxial pick-up coils.

The temperature dependence of the electric polarization along the $c$ axis ($P_c$) of Sr$_2$CoSi$_2$O$_7$ in a magnetic field of 8~T along various directions is shown in Fig.~1(b).
In the absence of the magnetic field, electric polarization did not emerge \cite{0jiba}.
When the magnetic field was applied along the [110] direction, a positive electric polarization emerged along the $c$ axis \cite{erratum}.
In \aa kermanite materials, electric polarization emerges even without poling electric fields, which is a unique feature of their multiferroicity \cite{Akaki,Murakawa}.
Therefore, the pyroelectric (displacement) current was measured, without a poling electric field.
By applying the magnetic field along the [1${\overline 1}$0] direction, the sign of the electric polarization reversed from the $+c$ direction, while the electric polarization was almost zero if the magnetic field was applied along the [100] direction. 
That is, an electric polarization reversal process was induced by a 90-degree rotation of the magnetic field.

Figure~2(a) shows the high magnetic field dependence of magnetization along the various axes of the Sr$_2$CoSi$_2$O$_7$ crystal at 1.4~K\@. 
When the magnetic field was applied along the [001] hard axis, the saturation field was 35~T with a saturation magnetization of about 3.3~$\mu_{\rm B}/{\rm Co}$. 
On the other hand, the saturation field was 17~T when the magnetic field was applied perpendicular to the $c$ axis, as reflected by the easy plane anisotropy.
By applying the magnetic field in the [100] direction, the saturation magnetization was about 3.2~$\mu_{\rm B}/{\rm Co}$. 
Contrary to the complete saturation in these directions, magnetization curves for fields parallel to the [110] direction show continuous increase above 17~T\@.

Figure~2(b) displays the electric polarization parallel to the $c$ axis of the Sr$_2$CoSi$_2$O$_7$ crystal as a function of external magnetic field applied in various directions at 1.4~K\@. 
When the magnetic field was applied along the [110] axis ($\phi=45^\circ$), the electric polarization rose rapidly in low fields, and reached a maximum at 8~T\@. 
With increasing fields further, the electric polarization gradually decreased and reversed sign at 16~T\@.
The result clearly demonstrates the existence of induced polarization even at 25~T, where all Co spins are forcibly aligned to the field direction as seen in Fig.~2(a).
As shown in the inset, although the azimuthal angle dependence alone can be derived from symmetry argument \cite{cryst}, we have to utilize a microscopic model to explain the complicated behavior of $P_c$ as a function of magnetic field.
In Ref.~13, the authors succeeded in reproducing this behavior with utilizing the {\it p-d} hybridization scenario for the canted antiferromagnetic spin system.
The similar calculation, however, is insufficient to explain the finite magnetoelectric effects persisting up to 55~T\@.
Recent theoretical calculation \cite{Penc} predicted the continuous change in $P_c$ and also the gradual increase in the magnetization before full saturation as a result of strong in-plane single ion anisotropy. 
This theory, however, predicts the similar gradual saturation of magnetization also for $H\parallel$ [100] that contradicts our results shown in Fig.~2(a).
In usual magnetic materials, finite slopes in the magnetization curves after spin saturation are ascribed to the contribution of the van Vleck effect, in which inclusion of the excited states having finite orbital angular momenta gives additional moments.
If we take into account of the anisotropic electron distribution in the excited state, and its anisotropic hybridization to the surrounding ligand orbitals, it can cause the additional electric polarization.
It is important to note that in the given crystal symmetry of Sr$_2$CoSi$_2$O$_7$, $P_c$ is proportional to $\sin 2\phi$ whatever the underlying mechanism is.

Figure~3 displays the temperature dependence of (a) magnetization in the [110] direction and (b) electric polarization along the $c$ axis of Sr$_2$CoSi$_2$O$_7$ crystal as a function of an external magnetic field parallel to the [110] direction. 
At 1.4~K below $T_{\rm N}$, since there is an antiferromagnetic long-range order of the Co spins, a large electric polarization is observed. 
On the other hand, the electric polarization can be sustained even at high temperature above $T_{\rm N}$ without magnetic order.
In this material, therefore, magnetic field-induced electric polarization is expected to be observed even at temperatures far above the magnetic transition.  
In fact, a small but finite electric polarization is discerned even at 300~K, although it is hard to see in Fig.~3(b).
The polarization data obtained for the paramagnetic phase at all temperatures are scaled with the square of the induced magnetization and nearly proportional to the square of the applied magnetic field. 
This behavior can in principle be explained as a magnetoelectric response of second order in the magnetic field, which is called the paramagnetoelectric effect \cite{ME}.
It is defined by the nonzero components of the tensor $\beta_{ijk}$:
\begin{equation}
P_i=\frac{1}{2}\beta_{ijk}H_jH_k.
\end{equation}
In the case of Sr$_2$CoSi$_2$O$_7$ with the point group $\overline{4}2m$, $\beta_{312}$ is nonzero.
Therefore, $P_c$ is most effectively induced by the applying magnetic field along the [110] direction.
According to Eq.~(1), the angular dependence of $P_c$ by applying the magnetic field $H_0$ with the azimuthal angle $\phi$ in the (001) plane will be proportional to $H_0^2\sin2\phi$.
This response supports our experimental results.

Next, the magnetoelectric response of Sr$_2$CoSi$_2$O$_7$ in low magnetic fields at room temperature is demonstrated. 
In order to realize a magnetic field-induced electric polarization at room temperature, electric polarization measurements were performed in a high-speed rotating magnetic field. 
By rotating the magnetic field at higher speeds, the electric polarization reversal process becomes faster, and the resultant displacement current becomes larger. 
Figure~4(d) shows the experimental configuration. 
The Nd-Fe-B permanent magnet was rotated around the sample by the motor to generate a rotating magnetic field within the (001) plane, and the displacement current flowing through the sample along the $c$ axis was observed. 
The [100] component of the rotating magnetic field was probed using a pick-up coil. 
Figure~4 shows the time dependence of (a) the [100] component of the rotating magnetic field, (b) the displacement current $I_c$, and (c) the electric polarization $P_c$ at room temperature.
The current $I_c$ can be decomposed into two components having the same and half of the rotation period of the magnetic field.
In Fig.~4(b), the observed current data are marked by dots, and the solid line shows the fitting sine curve having half the period of rotation of the magnetic field.
According to  Eq.~(1),  $P_c$ can be expressed as $P_c\propto \sin 2\phi$.
Consequently, the current component having half the rotation period of the magnetic field [solid line in Fig.~4(b)] arises from the intrinsic change of the electric polarization. 
On the other hand, a small additional component having the same rotation period as the magnetic field is attributed to measurement noise, such as would be due to an induced current flowing through a lead wire.
The time dependence of the electric polarization obtained by integration of the current component having the half period is shown in Fig.~4(c).
Thus, the magnetoelectric response in a low magnetic field has been successfully observed, even at room temperature. 
The phenomenon above, in which the current was induced by a change in the direction of the applied magnetic field, can be regarded as an effect similar to conventional electromagnetic induction.

In summary, the magnetic and dielectric properties of a Sr$_2$CoSi$_2$O$_7$ single crystal have been investigated in high magnetic fields.
When magnetization is saturated, the azimuthal angle $\phi$ dependence of the electric polarization at a fixed field of 25~T is proportional to $\sin 2\phi$.
Magnetic field-induced electric polarization is observed even in a paramagnetic state without magnetic order.
Although these behaviors can be regarded as conventional magnetoelectric effects of the second order, symmetry argument alone does not explain the observed complicated change in polarization as a function of magnetic field. 
From a microscopic point of view, the field-induced directional change in the hybridization of the Co $3d$ and O $2p$ orbitals can be relevant to these phenomena.
Our results in high magnetic fields suggest the importance of the orbital angular momenta in the excited states.
In addition, an intriguing magnetoelectric response has been demonstrated, with the low magnetic field-driven electric polarization flip, even at room temperature.
This new data opens up the potential to generate next-generation spintronic devices operating at room temperature with low power consumption.

We thank D. Akahoshi and M. Mochizuki for helpful discussions.
This work was partly supported by a Grant-in-Aid for JSPS Fellows from the Japan Society for the Promotion of Science, and also by the Ministry of Education, Culture, Sports, Science and Technology, Japan, through Grant-in-Aid for Scientific Research (23340096).

\clearpage
\begin{figure}
\begin{center}
\includegraphics[width=0.48\textwidth,clip]{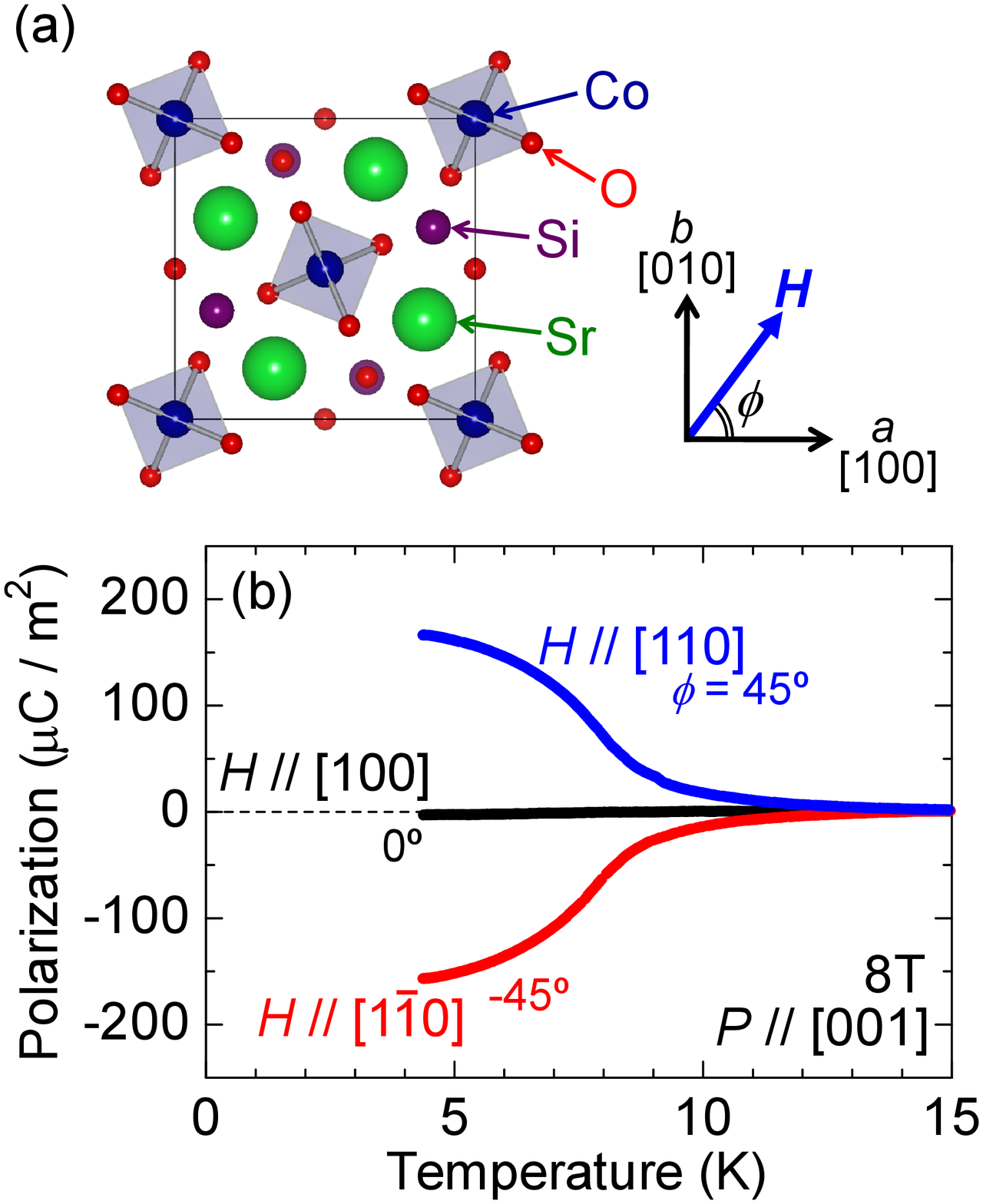}
\vspace{-9pt}
\caption{\label{fig1} (Color online) (a) Schematic crystal structure of Sr$_2$CoSi$_2$O$_7$ projected onto the $ab$ plane. (b) Temperature dependence of electric polarization along the $c$ axis of Sr$_2$CoSi$_2$O$_7$ in a magnetic field of 8~T applied along various directions.}
\end{center}
\end{figure}

\begin{figure}
\begin{center}
\includegraphics[width=0.48\textwidth,clip]{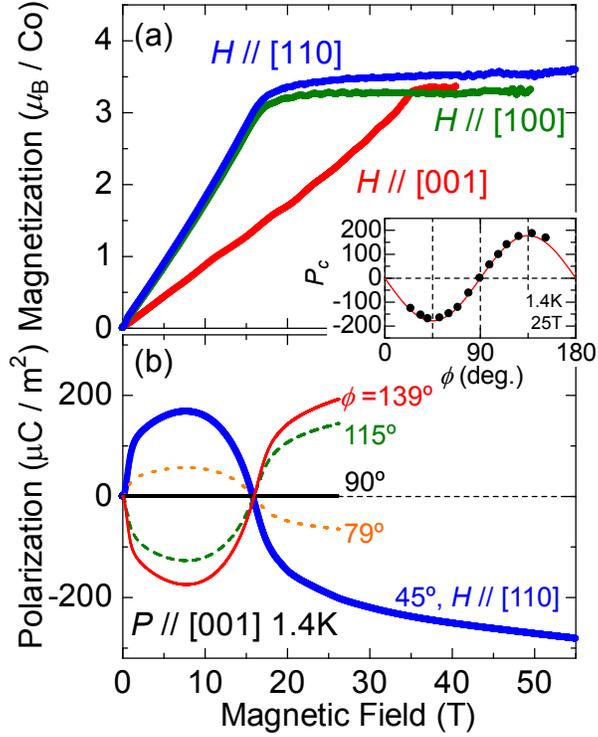}
\vspace{-9pt}
\caption{\label{fig2} (Color online) (a) Magnetization and (b) electric polarization along the $c$ axis of the Sr$_2$CoSi$_2$O$_7$ crystal as a function of external magnetic field applied along various directions at 1.4~K\@. The angle $\phi$ is defined as the angle between [100] and the $H$ direction in the (001) plane [see Fig.~1(a)]. The inset shows the angle $\phi$ dependence of the electric polarization at 25~T\@. The solid line represents $P_c(\phi)=P_c(45^\circ)\sin 2\phi$.}
\end{center}
\end{figure}

\begin{figure}
\begin{center}
\includegraphics[width=0.48\textwidth,clip]{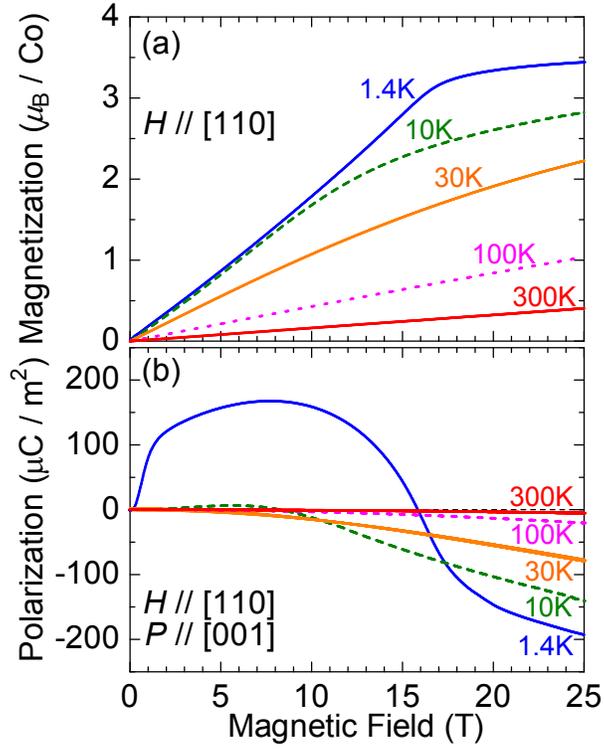}
\vspace{-9pt}
\caption{\label{fig3} (Color online) (a) Magnetization along the [110] direction and (b) electric polarization along the $c$ axis of the Sr$_2$CoSi$_2$O$_7$ crystal as a function of external magnetic field parallel to the [110] direction at several fixed temperatures.}
\end{center}
\end{figure}

\begin{figure}
\begin{center}
\includegraphics[width=0.48\textwidth,clip]{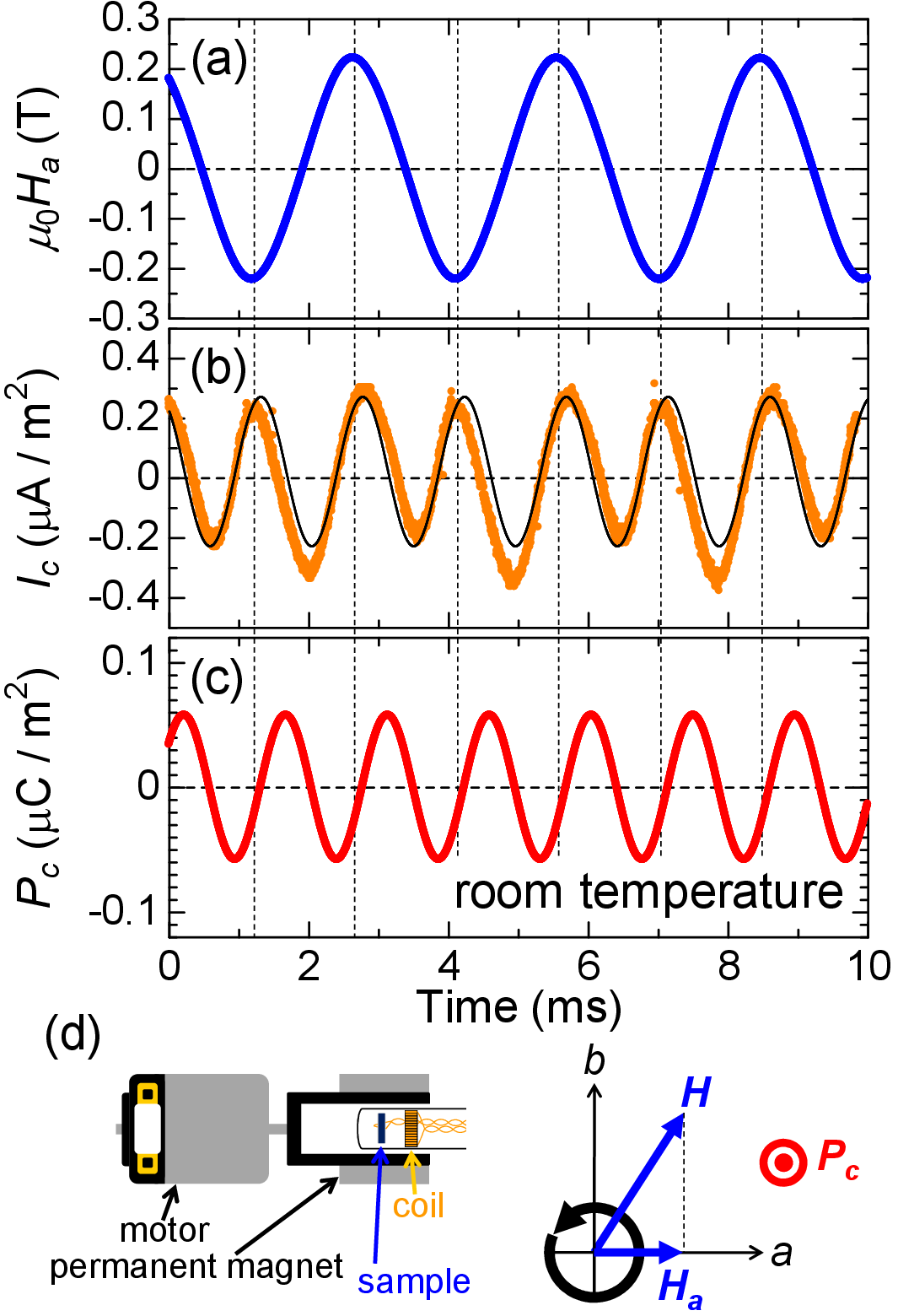}
\vspace{-9pt}
\caption{\label{fig4}  (Color online) Time dependence of (a) the $a$-axis component of the rotating magnetic field $H$, (b) the displacement current $I_c$, and (c) the electric polarization obtained along the $c$ axis at room temperature. (d) The experimental configuration of the electric polarization measurement in rotating $H$. The solid line in (b) represents a sine fit to the $I_c$ data (see text).}
\end{center}
\end{figure}


\begin{thebibliography}{99}
\bibitem{0jiba} Assuming that the sample forms a single magnetic domain, a finite electric polarization would emerge even in the absence of a magnetic field \cite{Murakawa}. However, macroscopic electric polarization at zero field does not appear due to the effect of quantum fluctuation \cite{Miyahara}.
\end{thebibliography}
\end{document}